\pdfoutput=1

\documentclass[10pt]{article}
\usepackage[letterpaper]{geometry}
\usepackage{hicss}
\usepackage{times}
\usepackage[none]{hyphenat}
\usepackage{url}
\usepackage{latexsym}
\usepackage{indentfirst}
\usepackage{graphicx}
\graphicspath{{Figures/}{images/}}

\usepackage[
    style=apa,
  ]{biblatex}

\usepackage{tabularx}
\usepackage{array,ragged2e}
\usepackage{booktabs}
\usepackage{multirow}
\usepackage[table]{xcolor}
\usepackage{float}
\usepackage{hyperref}
\usepackage{enumitem}
\usepackage{quoting}%
\usepackage{adjustbox}
\usepackage{tikz}
\usetikzlibrary{arrows.meta,positioning,fit,backgrounds,calc,shapes.symbols,shapes.arrows,shapes.misc}

\newcolumntype{P}[1]{>{\RaggedRight\arraybackslash}p{#1}}

\definecolor{uiomfp@light@blue}{RGB}{222,235,247}

\definecolor{mycolor}{RGB}{0, 0, 0}

\usetikzlibrary{arrows.meta,calc}

\DeclareDelimFormat[textcite]{finalnamedelim}{\addspace\bibstring{and}\space}

\newcommand{\org}{FinTech}

\title{Adopting Embedded AI Through an Innovation Team}

\author{
\begin{tabular}[t]{c}
Viktoria Stray\\
University of Oslo, SINTEF \\
\underline{stray@ifi.uio.no}
\end{tabular}
\hspace{0.21cm}
\begin{tabular}[t]{c}
Kristina Kantima \O stvold\\
University of Oslo\\
\underline{kriskos@ifi.uio.no}
\end{tabular}
\hspace{0.21cm}
\begin{tabular}[t]{c}
Agnes Hunstad Birkemo\\
University of Oslo \\
\underline{agneshb@ifi.uio.no}
\end{tabular}
\hspace{0.21cm}
\begin{tabular}[t]{c}
Nils Brede Moe\\
SINTEF\\
\underline{nils.b.moe@sintef.no}
\end{tabular}
}

\date{}

\begin{document}
\maketitle

\begin{abstract}

Software organizations are increasingly embedding AI capabilities into products that existing teams already own and maintain, but how this development work should be organized is not well understood. We conducted a qualitative case study of a temporary cross-functional innovation team in a large Norwegian financial organization, drawing on 18 interviews with 15 participants, 10 days of office observation, observations of ten meetings, and internal documents. We found that structural separation, AI competence inside the team, high autonomy within a clear mandate, senior expertise, and flexible ways of working enabled the team to explore and build embedded AI solutions. While structural separation enabled this work, it also made adoption dependent on coordination with the permanent product team that would own the solution. Misaligned roadmaps, prioritization gaps, and a lack of co-location constrained that coordination. We identify seven success factors for adopting embedded AI through an innovation team and propose expertise awareness as an alternative to redundancy in the Agile Team Effectiveness Model.

\end{abstract}

\subsubsection*{Keywords:}

AI adoption, self-organizing teams, empirical software engineering, agile development

\section{Introduction}
\label{sec:intro}
Organizations across industries are investing heavily in artificial intelligence (AI) and seek to embed AI capabilities into existing products and services \parencite{gronli2026introduction, stray_generative_2025}. However, successful AI adoption depends on access to specialized AI competence, which remains scarce in many organizations \parencite{abhari2026augmentation}. Rather than embedding AI specialists in every product team, organizations increasingly establish dedicated teams that explore AI opportunities and develop AI solutions for broader organizational use. While such teams may accelerate experimentation and innovation, they also create a challenge: AI solutions must eventually be adopted, maintained, and evolved by permanent product teams. How dedicated innovation teams can support embedded AI adoption in established organizations remains poorly understood.

The challenge is amplified because embedded AI development differs from traditional software development. AI development is data-dependent, uncertain, and experimental, making the work more exploratory and less predictable than conventional software engineering \parencite{giray_software_2021, amershi_se_ai, toniolo2026ai}. Developing software with embedded AI also requires specialized competence, typically held by data scientists. Integrating such competence into software development is not straightforward, as data scientists and software developers often bring different backgrounds, terminologies, and ways of working \parencite{hukkelbergExploringChallengesIntegrating2019, ulfsnesExploringOrganizationalModels2024}.

One response to these challenges is to organize AI expertise in dedicated innovation teams. Such teams resemble exploratory units described in the literature on organizational ambidexterity, where organizations separate exploratory work from operational product development \parencite{oreillyAmbidextrousOrganization2004}. Studies of cross-functional innovation teams \parencite{stippInnovationCrossfunctionalTeams2018} and Lean internal startups \parencite{edisonLeanInternalStartups2018} similarly highlight the value of dedicated teams that can experiment with new technologies while drawing on resources from the larger organization.

To address this gap, we conducted a qualitative case study at a large Norwegian financial organization. The organization established a dedicated innovation team to explore and develop embedded AI solutions while relying on existing product teams for long-term ownership and maintenance. This arrangement allowed us to examine both what enables a dedicated innovation team to develop embedded AI and what tensions arise when exploratory AI solutions are integrated into established product development structures. We ask:

    \vspace{-4pt}
\begin{description}
    \setlength{\itemsep}{2pt}
    \setlength{\parsep}{0pt}

    \item[\textbf{RQ1:}]
    What team characteristics enable an innovation team
    to develop embedded AI solutions in a large established organization?

    \item[\textbf{RQ2:}]
    What coordination tensions arise when such solutions must be adopted
    by the permanent product team that will take over ownership?
\end{description}
\vspace{-5pt}

This paper makes three contributions to research on software development. First, we provide an empirical account of how a temporary cross-functional innovation team developed embedded AI in a large established financial organization. Where research on collaboration between data scientists and software engineers has examined the integration of roles within a team \parencite{nahar2022collaboration,ulfsnesExploringOrganizationalModels2024}, we examine the relationship between two teams, where the team that holds the AI competence is not the team that will own the result. Second, we identify seven success factors for adopting embedded AI through an innovation team. Third, we extend agile team effectiveness theory by showing that redundancy may be a boundary condition in highly specialized AI teams, which can coordinate through expertise awareness instead. What is new is not the separation of exploration from exploitation, which the ambidexterity literature has described for two decades. It is what that separation costs when the exploratory team writes code into a system another team owns and maintains, and where those costs appear, such as in roadmap alignment and review capacity. Throughout, we argue that separation enables exploration, but integration is necessary for adoption.

\section{Background}
\label{sec:background}
\subsection{Embedded AI Development}
\label{sec:bg-embedded-ai}
We use the term \textit{embedded AI} for products where AI components are integrated into larger software systems. In such systems, an AI model, for example a large language model, is one component within a broader product that also includes user interfaces, backend logic, data pipelines, and integrations with existing systems. Unlike a stand-alone AI application, embedded AI must function within an existing product and workflow, so developing it means building the surrounding software and integrations, not just the model. Embedded AI development also differs from AI-assisted development. Recent studies of AI in software teams, including our own, largely examine how developers use tools such as GitHub Copilot and ChatGPT in their own work, and how these tools change individual workflows and pair programming \parencite{stray2025human,  stray2025generative}. There, the AI is a tool the team uses. In embedded AI development, it is part of what the team builds. 

\begin{table*}[t]
\centering
\caption{Overview of conducted interviews and observations.}
\label{tab:data-collection}

\small
\renewcommand{\arraystretch}{1.06}
\setlength{\tabcolsep}{3.5pt}

\begin{minipage}[t]{0.49\textwidth}
\vspace{0pt}
\centering

\begin{tabularx}{\linewidth}
{@{}l>{\raggedright\arraybackslash}Xr@{}}

\multicolumn{3}{c}{\textbf{Interviews}} \\[-1pt]
\toprule
\textbf{Informant ID} & \textbf{Team} & \textbf{Duration(s)} \\
\midrule
I1  & Team Innovate        & 45 min \\
I2  & Team Local           & 60 + 30 min \\
I3  & Team Local           & 60 + 30 min \\
I4  & AI \& Analytics Dep. & 55 min \\
I5  & Team Local           & 60 min \\
I6  & Team Innovate        & 65 min \\
I7  & Team Innovate        & 60 min \\
I8  & Team Innovate        & 60 + 60 min \\
I9  & Tech Dep.            & 60 min \\
I10 & Team Innovate        & 50 min \\
I11 & Team Innovate        & 60 min \\
I12 & Team Innovate        & 60 min \\
I13 & AI \& Analytics Dep. & 40 min \\
I14 & Team Local           & 60 min \\
I15 & AI \& Analytics Dep. & 50 min \\
\bottomrule
\end{tabularx}

\end{minipage}
\hfill
\begin{minipage}[t]{0.49\textwidth}
\vspace{0pt}
\centering

\begin{tabularx}{\linewidth}
{@{}l>{\raggedright\arraybackslash}Xr@{}}

\multicolumn{3}{c}{\textbf{Observations}} \\[-1pt]
\toprule
\textbf{Activity} & \textbf{Participants} & \textbf{Length} \\
\midrule
Workshop
    & Team Innovate 
    & 3.5 h \\
Stand-up meeting
    & Team Innovate
    & 30 min \\
Stand-up meeting
    & Team Innovate
    & 30 min \\
Technical planning
    & Team Innovate
    & 60 min \\
Stand-up meeting
    & Team Innovate
    & 30 min \\
Stand-up meeting
    & Team Innovate
    & 30 min \\
Showtime
    & Team Local
    & 40 min \\
Sync meeting
    & Both teams
    & 38 min \\
Stand-up meeting
    & Team Innovate
    & 27 min \\
Retrospective
    & Team Innovate
    & 60 min \\
Office observations
    & Team Innovate
    & 10 days \\
Office observations
    & Team Local
    & 10 days \\
\bottomrule
\end{tabularx}

\end{minipage}

\vspace{3pt}
\begin{minipage}{\textwidth}
\footnotesize

\textit{Note.} Multiple durations indicate follow-up interviews.%
\end{minipage}

\end{table*}
Developing AI-based systems also differs from traditional software development. \textcite{amershi_se_ai} report that managing data, model evaluation, and the non-deterministic behavior of models introduce difficulties that conventional software engineering processes do not anticipate. \textcite{giray_software_2021} similarly describe machine learning systems as harder to specify and test, with quality strongly dependent on data. Together, this work characterizes AI development as data-dependent, uncertain, and experimental. What is less examined is how these characteristics shape the way a team organizes embedded AI work inside an established organization, where the AI component must be fitted into existing products and processes.

\subsection{Autonomy and Coordination}
\label{sec:bg-crossfunctional}
Developing embedded AI requires combining software engineering, data science, domain expertise, and product development competencies in cross-functional teams that can make decisions without waiting for others.
Autonomous teams are characterized by a high degree of responsibility and authority over their work, including planning, task allocation, and decision-making, where team members coordinate tasks internally, adapt how work is carried out based on needs and competencies, and take collective responsibility for team effectiveness and outcomes within a flexible framework \parencite{highsmithAgileProjectManagement2009, guzzoTEAMSORGANIZATIONSRecent1996}. \textcite{patanakulAutonomousTeamsNew2008} found that autonomous structures are particularly effective under high technological uncertainty. Autonomy is not unbounded, however. \textcite{lassenius_taxonomy_largescale} caution that too much autonomy can undermine synchronization and predictability, because teams in large organizations do not deliver alone. This challenge becomes particularly salient when exploratory teams depend on permanent product teams for implementation and adoption. Embedding AI into products adds AI competence to the list of capabilities a cross-functional team needs. Integrating data scientists into agile teams is difficult, as the roles differ in background, terminology, and approach \parencite{hukkelbergExploringChallengesIntegrating2019, ulfsnesExploringOrganizationalModels2024}.

Team effectiveness models describe the internal conditions under which a team performs well. We use the Agile Team Effectiveness Model (ATEM), which \textcite{strode_teamwork_2022} developed for agile software teams and which specifies eight components: shared mental models, communication, mutual trust, shared leadership, redundancy, peer feedback, adaptability, and team orientation. Two features of the model are particularly relevant to our case. First, it defines redundancy as the ability of team members to substitute for one another, which presumes some overlap in what members can do. Second, it treats the single team as the unit of analysis, so inter-team dependencies fall outside its scope. We use ATEM to analyze the innovation team, and we return to both features in the discussion.

\subsection{Innovation Teams}
\label{sec:bg-innovation}
Established organizations must continue to run their existing business while also exploring new opportunities. \textcite{oreillyAmbidextrousOrganization2004} describe the ambidextrous organization, in which exploratory units are kept structurally separate from exploitative units so that each can follow the processes, structures, and culture that suit it. Separation lets exploratory work proceed without being constrained by the demands of ongoing operations, and it shields the exploitative units from the disruption of innovation work.

A recurring way to organize exploration is the dedicated cross-functional innovation team. Teams composed around members' expertise are more likely to produce products and processes previously unknown to the organization \parencite{stippInnovationCrossfunctionalTeams2018}. For such teams to succeed, prior work points to enablers such as top management support, permission to break existing rules, and a personal stake in the outcome, and to inhibitors such as dependencies on external teams \parencite{edisonLeanInternalStartups2018}. Experience matters too, with a positive relationship reported between experience and innovation performance \parencite{blindenbach-driessenInEffectivenessCrossFunctionalInnovation2015}. Together, this literature explains why a separated, senior, cross-functional team can be effective for exploration. It says less about what happens when the exploratory unit must hand its output to a permanent team for the innovation to be adopted.

\section{Method}
\label{sec:method}

We conducted a qualitative case study \parencite{yin_case_2009} of embedded AI development and AI adoption in \org{}, a large Norwegian organization in banking and insurance with roughly 2,300 employees. We studied Team Innovate, established to develop embedded AI solutions, together with Team Local, a permanent product team that Team Innovate depended on for domain knowledge and that would later own one of the solutions. Company and team names are pseudonyms. The study was carried out as part of a larger research project on AI adoption in software development, and the data were collected in connection with the master's thesis of the second and third authors \parencite{birkemo_ostvold_ai_2026}.%

\subsection{Data Collection}

We collected data from October 2025 to March 2026 
through interviews and observations.  We conducted 18 semi-structured interviews with 15 participants (see Table \ref{tab:data-collection}). The interviews lasted between 30 and 65 minutes, with an average of about 54 minutes. We interviewed all members of Team Innovate, a subset of Team Local who were involved in the shared project, and three people in management roles, covering two product managers, one tech lead, four developers, two designers, three data scientists, an agile coach, the senior vice president of the technology department, and the head of the analytics and AI department. Three interviews were follow-ups. We conducted 10 interviews in person and 8 over Microsoft Teams. Two or three researchers were present in each interview, with one leading the interview and the others observing and taking notes. All interviews were audio-recorded, transcribed automatically, and then corrected against the recordings to ensure accuracy.

We carried out 10 days of office observations in each of the teams to observe their daily work, ad hoc problem solving, and conversations. We also observed agile meetings (Table~\ref{tab:data-collection}) across both teams and a workshop in which Team Innovate taught other parts of the technology department how to build AI agents. Field notes followed a shared observation protocol. Our data collection ended before the planned handover of Project Alpha to Team Local. We studied the coordination that precedes adoption rather than the handover itself, which let us observe the dependencies between the two teams as the participants managed them from day to day. Our claims therefore concern the conditions for adoption and not its outcome.

\subsection{Data analysis} We analyzed the data using thematic analysis \parencite{braun_using_2006}, with a hybrid coding strategy. We coded in NVivo, beginning with an initial set of deductive codes informed by the Agile Team Effectiveness Model (ATEM) \parencite{strode_teamwork_2022}, and added inductive codes for themes that emerged beyond it. For example, \textit{redundancy} was a deductive code taken from the ATEM, while inductive codes such as \textit{Java vs. Python} and \textit{different pipelines} were grouped into a theme on the technology stack that we report in Section~\ref{sec:53}. We conducted the analysis in parallel with data collection, which let early findings shape the later interviews. Throughout, we wrote memos to record reflections and analytic decisions.%

\section{Case Context}
\label{sec:context}
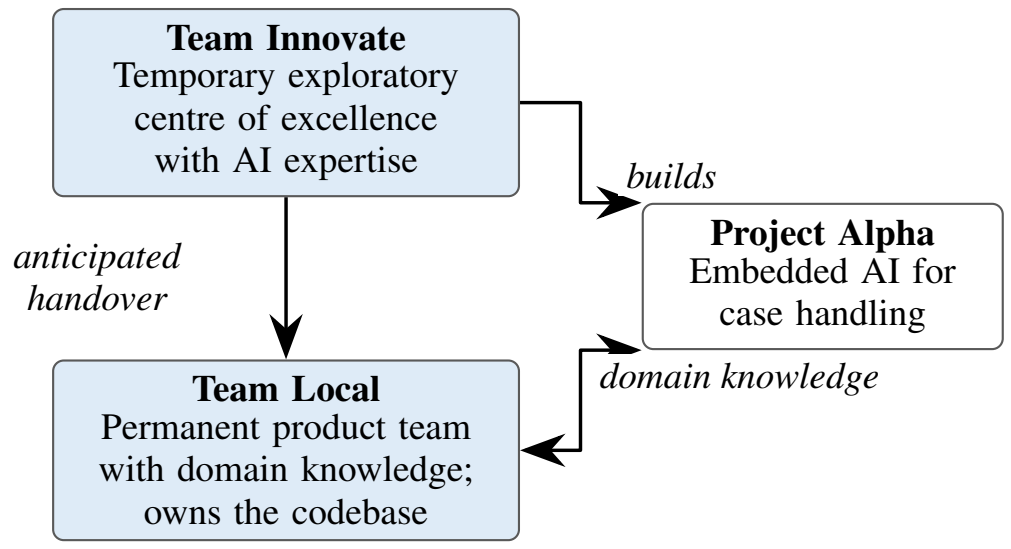
\begin{figure}[t]
\centering

\begin{tikzpicture}[
    font=\scriptsize,
    >={Stealth[length=3mm,width=2.2mm]},
    team/.style={
        draw=black!65,
        line width=0.4pt,
        rounded corners=2pt,
        fill=uiomfp@light@blue,
        align=center,
        text width=2.8cm,
        inner xsep=4pt,
        inner ysep=3pt
    },
    project/.style={
        draw=black!65,
        line width=0.4pt,
        rounded corners=2pt,
        fill=white,
        align=center,
        text width=2.1cm,
        inner xsep=4pt,
        inner ysep=3pt
    },
    arrow/.style={
        ->,
        draw=black,
        line width=0.65pt,
        line join=round
    },
    biarrow/.style={
        <->,
        draw=black,
        line width=0.65pt,
        line join=round
    },
    lbl/.style={
        font=\scriptsize\itshape,
        fill=white,
        inner sep=1.5pt,
        align=center
    }
]

\node[team] (innovate) at (0,1.15) {%
    \textbf{Team Innovate}\\[-1pt]
    Temporary exploratory\\
    centre of excellence\\
    with AI expertise};

\node[team] (local) at (0,-1.15) {%
    \textbf{Team Local}\\[-1pt]
    Permanent product team\\
    with domain knowledge;\\
    owns the codebase};

\node[project] (project) at (3.55,0) {%
    \textbf{Project Alpha}\\[-1pt]
    Embedded AI for\\
    case handling};

\draw[arrow]
    (innovate.east)
    -- ++(0.40,0)
    |- (project.north west);

\node[lbl] at (2.55,0.67) {builds};

\draw[arrow]
    (innovate.south)
    -- (local.north);

\node[lbl] at (-1.25,0) {anticipated\\handover};

\draw[biarrow]
    (local.east)
    -- ++(0.40,0)
    |- (project.south west);

\node[lbl] at (3.00,-0.68) {domain knowledge};

\end{tikzpicture}

\caption{Structurally separated teams in the embedded AI adoption case.}
\label{fig:case-overview}
\end{figure}

\org{}'s technology department is organized into cross-functional teams that typically combine developers, designers, product managers, and business representatives. \org{} had set a goal of adopting AI and increasing automation by embedding AI capabilities into its existing products. To pursue this, it established a temporary innovation team while the rest of the organization continued ordinary operations. Figure~\ref{fig:case-overview} gives an overview of the two teams we studied and their relationship.

\textbf{Team Local} is a permanent, cross-functional product team that develops and maintains systems for case handling in several insurance domains and holds the domain knowledge for these areas. The systems serve both internal users, such as case handlers, and external customers. The team has approximately ten members and follows the organization's Focus-Flex cadence, consisting of six-week Focus periods for prioritized delivery followed by two-week Flex periods for planning, feedback, and preparation. During our study, a data scientist was added to the team as a permanent member.

\textbf{Team Innovate} is a temporary, cross-functional team set up to explore and develop embedded AI solutions and to support \org{}'s wider goal of AI adoption. It functions as a center of excellence: beyond building specific products, it establishes practices for embedded AI development that other teams can reuse. The team has seven members: a product manager, a full-stack developer who is also the tech lead, a platform developer, a service designer, a prompt engineer, and two data scientists, and operates with a high degree of autonomy. It is co-located in a shared workspace and works in a flexible, self-defined rhythm rather than the organization's Focus-Flex cadence. Because it is intended to exist only temporarily, the solutions they build are handed over to permanent product teams that become their owners.

\textbf{Project Alpha.} During our study, Team Innovate was developing an embedded AI solution to support and partly automate a document-intensive case-handling process within Team Local's product domain. The end users were operational staff who handled incoming cases, and the AI model interpreted domain-specific documents inside an internal case-handling tool. The project was selected because the solution had the potential to scale to similar use cases and because the domain involved fewer regulatory constraints than more sensitive business areas, making it a feasible setting for experimentation.

Team Innovate held the AI expertise but depended on Team Local for domain knowledge, access to existing systems, and approval of pull requests in Team Local's codebase, while Team Local depended on Team Innovate for AI expertise. Project Alpha was full-time work for Team Innovate but sat outside Team Local's roadmap, handled alongside other responsibilities.%

\section{Findings}
\label{sec:findings}

We present five findings on how Team Innovate developed embedded AI: the work was exploratory, separation enabled that exploration, it required AI competence inside the team, internal dynamics supported experimentation, and adoption would depend on coordination with the permanent product team.

\subsection{Exploratory AI Development}
\label{sec:f0}
Team Innovate developed embedded AI through a sequence of phases that began with identifying an AI use case and understanding the domain, then moved through early experiments, a proof of concept, the design of a first version, and continuous iteration. The early phases were dominated by exploration and problem formulation rather than implementation. As one participant put it, much of the work was about understanding what to build before building it.

Being data-dependent, uncertain, and non-deterministic, the work required a flexible workflow that could be adjusted as new insights emerged, and it required data science competence inside the team, which meant collaboration across roles with different ways of working. The remaining findings show how Team Innovate's design responded to these conditions.

\subsection{Separation Enabled Exploration}
\label{sec:f1}
Team Innovate was deliberately set up as a temporary unit, structurally separate from the rest of the technology department and funded one year at a time. Separation was intentional. Participants explained that the team needed to remain free of ownership and maintenance duties to keep exploring. I8 put it directly: if the team had to own and maintain its solutions, ``\textit{the innovation curve}'' would ``\textit{flatten out}'', and the team could no longer serve its purpose.

Within this separated unit, autonomy was the central design choice. The two people who established the team described giving it a ``\textit{very broad assignment, with a high degree of freedom to find the best way to solve it}'' [I13]. Management set a concrete goal but left the tools, approach, and rhythm to the team. The team members experienced this autonomy in their daily work. I10 described being ``\textit{highly autonomous in the sense that no one above us tells us what to do}''. A manager reflected, \textit{"This was part of the key to success. That they were allowed to operate with complete freedom."} [I13]
Autonomy was not unconditional. Both management and members described it as autonomy held to a high bar, where the team was expected to use its freedom responsibly.

\subsection{AI Competence Moved Into the Team}
\label{sec:53}

AI competence moved from a central analytics unit into the team that built the product. Traditionally, data scientists at \org{} worked outside the agile software teams, acting as what participants called ``internal consultants''. In Team Innovate, the data scientist was instead a full member of the cross-functional team, and I6 described the change as a model where ``a data scientist is embedded in a team''.

Embedding the competence mattered because dependencies on other teams were the main source of delay. I12 described this as the ``\textit{biggest blocker}'' in a large organization: ``\textit{you often have to wait for someone else to fix something for you. It often takes a long time, and you end up waiting. So the fact that we could own the entire product end-to-end made things much easier}''. The dependency ran the other way, too. A member of Team Local recalled trying to build an AI agent during a hackathon and finding that ``\textit{we didn't quite know how to tune it. That's when we realized how dependent we are on the AI people}'' [I3].

Integrating the two roles was not friction-free. Participants described the gap between data scientists and developers as cultural, and it showed most clearly in two places. The roles judged results differently: developers treated a task as done once data moved through the system, while data scientists saw that point as the start of assessing quality, asking whether a result was ``good enough'' at 95 percent rather than true or false. The sharper divide was the technology stack. Eleven of fifteen participants noted that data scientists worked in Python while FinTech's systems ran on Java, and because the organization's repositories, security, and standards existed only for Java, Python solutions did not meet production requirements and had to be rewritten. As I2 put it, ``\textit{we have a data science team that writes everything in Python [.\!.\!.] but when we try to use it, we have very rigid guidelines saying that we are not allowed to use Python}'' [I2], so the Python solutions had to be translated to Java. I4 gave a concrete example of an embedded AI solution that was ``\textit{first fully developed in Python and ready for production. But then it was reviewed and rejected, because it wasn't considered good enough. [.\!.\!.] The code then had to be translated into Java}'' [I4]. The team absorbed the cultural differences through co-location and daily collaboration, but the tech-stack divide reflected organizational standards it could not change on its own.

The same move was underway in Team Local, which added a data scientist as a permanent member during our study. A manager described this as a deliberate shift toward ``\textit{end-to-end ownership of the solutions}'', prompted by the realization that ``\textit{the transition from having a pilot that works on someone's desktop to something that's actually integrated into the system involves a lot of dependencies to other teams}'' [I15]. That both teams moved the same way suggests the relocation of AI competence into the product team was becoming a general pattern at FinTech, not a feature of Team Innovate alone.
\begin{table*}[t]
    \centering
    \small
    \caption{How the Agile Team Effectiveness Model (ATEM) components appeared in Team Innovate.}
    \label{tab:atem-team-innovate}

    \setlength{\tabcolsep}{4pt}
    \renewcommand{\arraystretch}{1.04}

    \begin{tabularx}{\textwidth}{
        >{\raggedright\arraybackslash}p{2.0cm}
        >{\raggedright\arraybackslash}p{4.1cm}
        >{\raggedright\arraybackslash}X}

        \toprule
        \rowcolor{uiomfp@light@blue}
        \textbf{Component} &
        \textbf{Manifestation} &
        \textbf{Illustrative quote} \\
        \midrule

        Shared mental models
            & Shared understanding of goals, product, and expertise
            & ``There's a lot of discussion about how we do things, why we do things [...] that's an investment that we're trying to make in order for all of us to have the same mental model.'' [I11] \\

        \addlinespace[1pt]

        Communication
            & Meetings plus informal co-located communication
            & ``And then we usually sit together [...] What I think has worked very well is that we are not far from each other.'' [I7] \\

        \addlinespace[1pt]

        Mutual trust
            & High trust grounded in expertise
            & ``It's trust, but it's trust based on experience.'' [I8] \\

        \addlinespace[1pt]

        Shared leadership
            & Decentralised, competence-based decisions
            & ``We are all equally weighted because we all cover a different aspect of a product.'' [I8] \\

        \addlinespace[1pt]

        Redundancy
            & Low role overlap; shared understanding
            & ``There isn't really any exchange of experience in that sense, because it is such a highly specialised team.'' [I10] \\

        \addlinespace[1pt]

        Peer feedback
            & Frequent formal and informal feedback
            & ``So we also would give feedback [...] when things can be improved [...] And then we try to reason through these discussions together.'' [I11] \\

        \addlinespace[1pt]

        Adaptability
            & Continuous adjustment of solutions and practices
            & ``Should we build one ourselves, or should we lean more towards the platform? And then we ended up scrapping the one we built ourselves, and going all in on [the platform].'' [I1] \\

        \addlinespace[1pt]

        Team orientation
            & Shared goals, joint decisions, high cohesion
            & ``This team is the best. We share the same goals [...] We don't always agree, but we can always talk about it. I don't think we could have done it without being the people that we are.'' [I10] \\

        \bottomrule
    \end{tabularx}
\end{table*}
\subsection{Team Dynamics and the ATEM}
\label{sec:f3}
The team's autonomy worked because of how the members collaborated, and we describe this collaboration through components of ATEM \parencite{strode_teamwork_2022}. \textit{Shared leadership} was grounded in expertise rather than position. The product manager acted as a servant leader, handling boundary-spanning work and removing blockers, while decisions were resolved by whoever held the relevant knowledge. As I12 put it, ``\textit{everyone takes part in designing the system, but someone still has the authority to make the final decision once we disagree}''. \textit{Trust }in the team appeared to be rooted in the confidence in each other’s competence and experience within their respective domains. I1 described the team environment as open and characterized by an “implicit trust” that the team would find a way to reach its shared goals.

\textit{Redundancy }was limited, as senior specialists could not easily take over one another's tasks. \textit{Shared mental models} and co-location compensated. We observed that even during focused individual work, members made brief clarifications across desks and gathered around a screen or whiteboard to work through problems together. \textit{Communication} was calibrated to the dependencies between roles: members did not need to know everything, but had to know their own area well enough to make their work accessible to others [I6]. Members saw the value less in the code than in building a shared understanding of what they were building [I11]. \textit{Peer feedback } was observed in both structured meetings and informal discussions, where team members raised concerns, suggested improvements, and refined solutions together.

\textit{Adaptability }was central to how the team worked. It adjusted its practices, its meetings, and even its own composition as the project required, for example, shifting from separate frontend and backend developers to a single full-stack role. \textit{Team orientation} held this together. Members were senior, self-directed, and motivated by exploratory work, which let the team operate without close follow-up: ``\textit{we had everything we needed within the team, so there were not many external dependencies}'' [I12]. 
Team Innovate therefore displayed all eight components of the ATEM, with the exception of redundancy, which we return to in Section 6.1. Table~\ref{tab:atem-team-innovate} gives an illustrative quote for each component.

\subsection{Coordination Tensions Before Handover}
\label{sec:f4}
Team Innovate could explore and build on its own, but the solution could only be adopted if Team Local understood it, prioritized it, and eventually owned it. The two teams depended on each other: Team Innovate held the AI expertise and the mandate to explore; Team Local held the domain knowledge, the existing systems, and the codebase. Managing this dependency required coordination, and at the time of our study that coordination was still at an early stage. After initial conversations about the domain, Team Innovate worked largely on its own, partly because the solution was, as I11 put it, ``\textit{not mature enough yet to have proper collaboration}'', and partly because the team's goal reached beyond the specific insurance case: ``\textit{our goal is a bit bigger than that. It is the technology as a whole. So yes, we are going to involve Team Local more at some time}'' [I12].

The teams had nonetheless established a set of coordination mechanisms: a shared Slack channel, a shared GitHub workspace where Team Innovate contributed pull requests to Team Local's services, workshops, and status meetings between the product managers. The PM meetings were intended weekly but held roughly every other week, and workshop attendance from Team Local was optional and competed with their own work.

This coordination was constrained by a gap in priorities. Project Alpha was Team Innovate's full-time work, but sat outside Team Local's roadmap. I14 described it plainly: ``\textit{It's not driven by our team, it's driven by Team Innovate. So it's not in our roadmap. It's listed as something we're involved in}''. Team Local engaged with the project alongside other work, and as I2 put it, ``\textit{we don't work dedicated one hundred percent with AI. If we're lucky, we might get to work with it ten percent, on the side}''. I12 linked the gap to the difficulty of collaborating well: \textit{"They have their own backlog with their own problems, and then we come in from the side with a very specific problem that they might not have time to look at. So it becomes a challenge."} [I12]

The gap surfaced as concrete friction. We observed delays in pull-request reviews repeatedly raised as a source of frustration in Team Innovate, with one member blocked while waiting for approval from Team Local. The situation improved after the product manager raised it in a stand-up, but the dependence on a team with different priorities slowed progress. In a retrospective, members wrote that the ``\textit{collaboration with the local team was difficult}'', that they were ``\textit{waiting too long for pull request from local team}'', and that the collaboration ``\textit{felt a bit like a cold shoulder sometimes}''.

Coordination at the developer level was thin, and the PM status meetings did not fill the gap. Team Local would eventually own the solution but was not closely involved in building it, and changes to its systems could arrive without warning. I14 described one such case, where an AI component was connected to an application already in daily use: \textit{''It felt like they had gone into our application and made changes. The application we built is used daily by advisors, and they noticed disruptions [.\!.\!.] It wasn't enough with the PM status meetings. The developers themselves hadn't really communicated directly, and that became a disadvantage.''} [I14]

A lack of co-location widened the priority gap, and what we observed cuts against the obvious remedy. The teams sat apart, so interaction depended on digital channels and infrequent meetings. I14 saw this as an obstacle: ``\textit{We work too far apart from each other. We're separated.}'' I12 described the rhythm, ``\textit{we have a meeting there, and then three weeks pass by, and then you have another meeting}'', and suggested ``\textit{that everyone sits together every Tuesday}''. Yet physical proximity on its own changed little. When Team Local moved to the floor below during our study, I14 reported no difference: ``\textit{Whether it's another floor or another city, it's basically the same}''. What mattered was whether the specific people a piece of work depended on sat together, not the building-wide distance.

Team Innovate treated the relationship with Team Local as something to be managed rather than assumed. Entering another team's domain was delicate. As I13 put it, ``\textit{now that we entered a space where there is a very clear ownership, it has been more challenging [...] When you try to innovate on something, you will inevitably step on someone's toes}''. The team responded with two deliberate practices: positioning and transparency.

The team positioned itself as a guest in Team Local's domain rather than an expert taking it over. I8 made this explicit: ``\textit{you guys are the local domain experts. This is your day-to-day. We're just tourists and we've got this idea where we think we can help out}'', and tied effective collaboration to soft skills: ``\textit{So soft skills in collaborating between teams is so important}''. The same care was extended to how the team called meetings and ran workshops, which it framed as requests for help rather than announcements.

The team also worked to stay transparent about what it was building. I11 described sharing plans before implementation so that Team Local was ``\textit{up-to-date and there are no big surprises}'', which gave the receiving team a chance to raise concerns before changes were made. I11 saw transparency as a way of managing the dynamic the separation created: ``\textit{insurance is their domain, and we're maybe flying from the side trying to help [...] But we are trying to be very transparent and very open. So we are sharing a lot of what we do ahead of doing it}''. We argue that positioning and transparency reduced the friction of entering another team's domain, but that these were practices internal to Team Innovate. They could not, on their own, close the structural gaps in roadmap, capacity, and co-location that constrained adoption.

\section{Discussion}
\label{sec:discussion}

\begin{table*}[t]
    \centering
    \small
    \caption{Success factors for adopting embedded AI through an innovation team.}
    \label{tab:success-factors}
    \setlength{\tabcolsep}{5pt}
    \renewcommand{\arraystretch}{1.25}
    \begin{tabularx}{\textwidth}{>{\raggedright\arraybackslash}p{4.2cm} >{\raggedright\arraybackslash}X}
        \toprule
        \rowcolor{uiomfp@light@blue}
        \textbf{Success factor} & \textbf{Description} \\
        \midrule
        1. Establish a structurally separate innovation team
            & Shield exploratory work from the routines and delivery demands of the permanent product teams, so it can develop with greater flexibility without disrupting ongoing operations. \\
        \addlinespace[3pt]
        2. Grant high autonomy within a clear mandate
            & Let the team choose its own tools, processes, and rhythm under a concrete goal, so it can adapt as its understanding of an uncertain, data-dependent problem evolves. \\
        \addlinespace[3pt]
        3. Compose a senior, cross-functional team
            & Staff the team with experienced members who together hold the competence to work end-to-end, reducing external dependencies and enabling independent decisions. \\
        \addlinespace[3pt]
        4. Bring AI competence into the team
            & Embed data scientists as team members rather than treating them as internal consultants, so AI competence is a capability of the team and not an external dependency. \\
        \addlinespace[3pt]
        5. Select people with intrinsic motivation
            & Choose members who are motivated by exploratory, open-ended work and a high degree of freedom rather than overwhelmed by it. \\
        \addlinespace[3pt]
        6. Co-locate the innovation team
            & Enable informal communication and spontaneous problem-solving so the team can build and maintain a shared understanding across specialised roles with limited redundancy. \\
        \addlinespace[3pt]
        7. Coordinate with the receiving product team
            & Align roadmaps and priorities with the permanent product team, ensure it has capacity to participate during critical phases, and seat the specific people who depend on each other together ahead of handover. \\
        \bottomrule
    \end{tabularx}
\end{table*}

\subsection{Separation Enables Exploration}
\label{sec:disc-separation}
Team Innovate was set up as an exploratory unit kept structurally separate from the exploitative units that run \org{}'s existing products. This mirrors the ambidextrous organization, in which exploratory and exploitative units are separated so that each can follow its own processes and culture \parencite{oreillyAmbidextrousOrganization2004}. Separation allowed Team Innovate to work in a flexible rhythm instead of the organization's Focus-Flex cadence, and it shielded the product teams from the disruption of exploratory work. We argue that this separation was a precondition for the team's exploratory embedded AI work because that work was uncertain, data-dependent, and hard to plan against the delivery routines of an ordinary product team.

Separation worked because of what was placed inside the separated unit. The team had high autonomy within a clear mandate, which matches evidence that autonomous structures are effective under high technological uncertainty \parencite{patanakulAutonomousTeamsNew2008}, and that team autonomy has to be adjusted to its organizational context when digital transformation increases the dependencies between teams \parencite{ravn_team_autonomy}. Autonomy on its own can overwhelm a team \parencite{grassEmpowermentDynamicsTeam2020}, but \org{} paired it with senior, cross-functional members who were intrinsically motivated by an exploratory mandate. This combination reflects findings that experience supports innovation performance \parencite{blindenbach-driessenInEffectivenessCrossFunctionalInnovation2015}, that expertise-based cross-functional teams produce genuinely new products \parencite{stippInnovationCrossfunctionalTeams2018}, and that a personal stake in the outcome enables internal startups \parencite{edisonLeanInternalStartups2018}. 

Embedding AI competence inside the team, instead of treating data scientists as internal consultants, removed the external dependency that participants identified as their biggest source of delay. This reflects a shift from a consultant-based enabling model to a team-member model for organizing data scientists \parencite{ulfsnesExploringOrganizationalModels2024}, and it changes what cross-functionality means: AI competence becomes part of the competence an agile team must hold to build embedded AI. Integrating data scientists is not friction-free, as they bring different backgrounds, terminologies, and ways of working \parencite{hukkelbergExploringChallengesIntegrating2019}, but the team's senior, cross-functional composition absorbed these differences.

Internally, the team displayed the dynamics that the Agile Team Effectiveness Model (ATEM) associates with effective teamwork, summarised in Table~\ref{tab:atem-team-innovate}. One notable exception was redundancy. The model treats redundancy as an important enabler of team effectiveness because members can substitute for one another when needed. In our case, however, such substitution was largely impossible due to the specialised expertise required for embedded AI development. Nevertheless, the team remained effective through a shared understanding of expertise distribution and competence-based leadership. This finding suggests that redundancy may not be a necessary condition for effectiveness in highly specialised innovation teams. We therefore propose that future versions of the ATEM incorporate expertise awareness as an alternative coordination mechanism alongside redundancy.

\subsection{Integration is Necessary for Adoption}
\label{sec:disc-integration}
The same separation that enabled exploration created the central problem for adoption. Team Innovate was not meant to own the solutions they developed, so adoption depended on Team Local's understanding, prioritizing, and taking over the work. Adoption was a problem of inter-team coordination, that is, of managing the dependencies between the two teams. Team Innovate depended on Team Local for domain knowledge, system access, and pull-request approval, and Team Local would inherit a solution it had not built. This kind of dependency falls outside common models of team effectiveness, which treat the team as the unit of analysis and exclude inter-team dependencies from their scope \parencite{strode_teamwork_2022}. Our case shows that for an innovation team whose output must be adopted elsewhere, the relationship to the receiving team is not peripheral. Adoption has to pass through it, and it was already under strain well before the planned handover.

Coordination between the two teams was constrained by misaligned roadmaps, a gap in prioritized capacity, and a lack of co-location. Project Alpha was full-time work for Team Innovate but sat outside Team Local's roadmap, so it competed with Team Local's own backlog. The mechanisms the teams had in place, a shared Slack channel, shared pull requests, workshops, and product-manager status meetings, were not sufficient on their own. They coordinated the managers more than the developers, so changes to Team Local's systems could arrive without warning and disrupt applications already in daily use. We argue that effective coordination here depended on structural conditions the innovation team could not create on its own, namely, roadmap alignment, sufficient capacity in the receiving team, and proximity between the specific developers the work depended on.

We expect our findings to hold where three features of our case are present. First, the AI component was built into a system owned and maintained by another team. Second, the receiving team operated under platform and security standards they could not change; for example, Python solutions had to be rewritten in Java. Third, the innovation team was temporary by design and could not own what it built.%

\subsection{Implications for Practice}
\label{sec:disc-practice}
For organizations seeking to embed AI into existing products through an innovation team, our findings translate into a set of success factors, shown in Table~\ref{tab:success-factors}. The factors fall into two groups: separate the team to enable exploration, then deliberately design its integration with the receiving product team. We argue that embedded AI adoption succeeds when organizations treat the relationship between the temporary innovation team and the permanent product team as something to be designed, not assumed.

\subsection{Limitations}

We assess validity following \textcite{runeson2009guidelines}.
\textbf{Construct validity.} Our constructs come from an established model of agile team effectiveness \parencite{strode_teamwork_2022}, and we triangulated interviews with 10 days of parallel office observation across both teams (20 researcher-days in total) and internal documents.
\textbf{Internal validity.} Access was arranged through a key informant in Team Innovate, which risks privileging that team's account, so we interviewed across both teams, three management levels, and the analytics department, and checked statements about one team against the other.
\textbf{External validity.} This is a single case in one organization and one sector, and we make no claim to statistical generalization. We offer the success factors as analytically generalizable propositions \parencite{yin_case_2009} and state the conditions under which we expect them to hold in Section~\ref{sec:disc-integration}.
\textbf{Reliability.} We coded in NVivo using a codebook that combined deductive ATEM codes with inductive codes. Our data ended before the planned handover, so our claims concern the conditions for adoption rather than its outcome.

\label{sec:disc-theory}

\section{Conclusion}
\label{sec:conclusion}
We studied a temporary cross-functional innovation team developing embedded AI in a large Norwegian financial organization, together with the permanent product team that would later own the solution. We conducted 18 interviews with 15 participants, carried out ten days of office observation in each of the teams, and observed ten meetings.

We found that developing embedded AI was exploratory, data-dependent, and difficult to plan for within ordinary delivery routines, and that it required AI competence within the team. Data scientists moved from internal consultants to full members of the team, which reduced external dependencies and supported end-to-end work. The innovation team was a separate, autonomous unit of senior members motivated by exploratory work, and it showed the dynamics that models of agile team effectiveness associate with success. Shared leadership and a shared understanding of who held which expertise compensated for its limited redundancy. The same separation that enabled this work also made adoption depend on coordination with the permanent product team, which was constrained by misaligned roadmaps, prioritization gaps, and a lack of co-location.

Future research could study the handover itself, test whether the success factors hold for other AI innovation teams, examine the mechanisms that align roadmaps and capacity across exploratory and exploitative units, and develop ways of working that join developers and data scientists in embedded AI development.

\section*{Acknowledgments}
We thank \org{} and all participants, who welcomed us into their daily work and openly shared their experiences. This research was supported by the Research Council of Norway through the Kairos project (grant 357147).

\printbibliography

@article{stray_generative_2025,
  author  = {Stray, Viktoria and Hanssen, Geir Kjetil and Barbala, Astri and Smite, Darja and Stol, Klaas-Jan},
  title   = {What is generative {AI} good for? {Introduction} to the special issue on generative {AI} in software engineering},
  journal = {Information and Software Technology},
  volume  = {187},
  pages   = {107857},
  year    = {2025},
  doi     = {10.1016/j.infsof.2025.107857}
}

@article{runeson2009guidelines,
  title={Guidelines for conducting and reporting case study research in software engineering},
  author={Runeson, Per and H{\"o}st, Martin},
  journal={Empirical software engineering},
  volume={14},
  number={2},
  pages={131--164},
  year={2009},
  publisher={Springer}
}

@inproceedings{nahar2022collaboration,
  title={Collaboration challenges in building ml-enabled systems: Communication, documentation, engineering, and process},
  author={Nahar, Nadia and Zhou, Shurui and Lewis, Grace and K{\"a}stner, Christian},
  booktitle={Proceedings of the 44th international conference on software engineering},
  pages={413--425},
  year={2022}
}

@inproceedings{toniolo2026ai,
  author    = {Toniolo, Korinzia},
  title     = {AI in Action: Toward a Theory of Hype Translation in Organizations},
  booktitle = {Proceedings of the 59th Hawaii International Conference on System Sciences},
  year      = {2026},
  pages     = {4011--4020},
  isbn      = {978-0-9981331-9-5},

}

@inproceedings{abhari2026augmentation,
  author    = {Abhari, Kaveh and Safaei Pour, Morteza and Sanatizadeh, Aida},
  title     = {Augmentation Innovation Paradox: Rethinking Validation in Generative AI-Driven Open Innovation},
  booktitle = {Proceedings of the 59th Hawaii International Conference on System Sciences},
  year      = {2026},
  pages     = {331--340},
  isbn      = {978-0-9981331-9-5}
}

@mastersthesis{birkemo_ostvold_ai_2026,
  author    = {Birkemo, Agnes Hunstad and {\O}stvold, Kristina Kantima},
  title     = {Adopting {AI} Through an Innovation Team:
               A Case Study of an Innovation Team in a Large Financial Organisation},
  school    = {University of Oslo},
  year      = {2026},
  type      = {Master's thesis},
  address   = {Oslo, Norway}
}

@book{highsmithAgileProjectManagement2009,
    title = {Agile {Project} {Management}: {Creating} {Innovative} {Products}},
    isbn = {978-0-321-65917-0},
    shorttitle = {Agile {Project} {Management}},
    language = {en},
    publisher = {Pearson Education},
    author = {Highsmith, Jim},
    month = jul,
    year = {2009},
}

@book{yin_case_2009,
    address = {Thousand Oaks, Calif},
    edition = {4th ed.},
    series = {Applied social research methods series},
    title = {Case study research: design and methods},
    volume = {vol. 5},
    isbn = {978-1-4129-6099-1},
    shorttitle = {Case study research},
    language = {eng},
    publisher = {Sage},
    author = {Yin, Robert K.},
    year = {2009},
}

@article{guzzoTEAMSORGANIZATIONSRecent1996,
    title = {Teams in organizations: {Recent} {Research} on {Performance} and {Effectiveness}},
    volume = {47},
    issn = {0066-4308, 1545-2085},
    language = {en},
    number = {Volume 47, 1996},
    urldate = {2026-05-03},
    journal = {Annual Review of Psychology},
    publisher = {Annual Reviews},
    author = {Guzzo, Richard A. and Dickson, Marcus W.},
    month = feb,
    year = {1996},
    pages = {307--338},
}

@inproceedings{gronli2026introduction,
  author    = {Gr{\o}nli, Tor-Morten and Kaindl, Hermann and Majchrzak, Tim A.},
  title     = {Introduction to the Minitrack on Software Technology and Software Development},
  booktitle = {Proceedings of the 59th Hawaii International Conference on System Sciences},
  year      = {2026},
  pages     = {7391--7392},
  isbn      = {978-0-9981331-9-5},
}

@article{oreillyAmbidextrousOrganization2004,
    title = {The ambidextrous organization.},
    journal = {Harvard business review},
    author = {O'Reilly, C. and Tushman, M.},
    month = apr,
    volume={82},
number={4},
    year = {2004},
}

@article{stippInnovationCrossfunctionalTeams2018,
    title = {Innovation and cross-functional teams: {Analysis} of innovative initiatives in a {Brazilian} public organization},
    volume = {24},
    issn = {1352-7592},
    shorttitle = {Innovation and cross-functional teams},
    number = {1-2},
    urldate = {2026-03-27},
    journal = {Team Performance Management},
    author = {Stipp, Daniel Marasquini and Pimenta, Márcio Lopes and Jugend, Daniel},
    month = mar,
    year = {2018},
    pages = {84--105},
}

@article{edisonLeanInternalStartups2018,
    title = {Lean {Internal} {Startups} for {Software} {Product} {Innovation} in {Large} {Companies}: {Enablers} and {Inhibitors}},
    volume = {135},
    issn = {0164-1212},
    shorttitle = {Lean {Internal} {Startups} for {Software} {Product} {Innovation} in {Large} {Companies}},
    url = {https://www.sciencedirect.com/science/article/pii/S0164121217302157},
    doi = {10.1016/j.jss.2017.09.034},
    urldate = {2026-03-29},
    journal = {Journal of Systems and Software},
    author = {Edison, Henry and Smørsgård, Nina M. and Wang, Xiaofeng and Abrahamsson, Pekka},
    month = jan,
    year = {2018},
    pages = {69--87},
}

@article{blindenbach-driessenInEffectivenessCrossFunctionalInnovation2015,
    title = {The ({In}){Effectiveness} of {Cross}-{Functional} {Innovation} {Teams}: {The} {Moderating} {Role} of {Organizational} {Context}},
    volume = {62},
    shorttitle = {The ({In}){Effectiveness} of {Cross}-{Functional} {Innovation} {Teams}},
    journal = {IEEE Transactions on Engineering Management},
    author = {Blindenbach-Driessen, Floortje},
    month = feb,
    year = {2015},
    pages = {29},
}

@inproceedings{patanakulAutonomousTeamsNew2008,
    title = {Autonomous teams and new product development},
    urldate = {2026-03-25},
    booktitle = {{PICMET} '08 - 2008 {Portland} {International} {Conference} on {Management} of {Engineering} \& {Technology}},
    author = {Patanakul, Peerasit and Chen, Jiyao and Lynn, Gary S.},
    month = jul,
    year = {2008},
    pages = {1214--1223},
}

@article{grassEmpowermentDynamicsTeam2020,
  title={From empowerment dynamics to team adaptability: Exploring and conceptualizing the continuous agile team innovation process},
  author={Grass, Anastasia and Backmann, Julia and Hoegl, Martin},
  journal={Journal of Product Innovation Management},
  volume={37},
  number={4},
  pages={324--351},
  year={2020},
  publisher={Wiley Online Library}
}

@INPROCEEDINGS{lassenius_taxonomy_largescale,
  author={Lassenius, Casper and Dingsøyr, Torgeir},
  booktitle={2025 IEEE/ACM 18th International Conference on Cooperative and Human Aspects of Software Engineering (CHASE)}, 
  title={Towards a Taxonomy for Autonomy in Large-Scale Agile Software Development}, 
  year={2025},
  volume={},
  number={},
  pages={96-101}
  }

@article{strode_teamwork_2022,
    title = {A teamwork effectiveness model for agile software development},
    volume = {27},
    issn = {1382-3256, 1573-7616},
    language = {en},
    number = {2},
    urldate = {2026-01-06},
    journal = {Empirical Software Engineering},
    author = {Strode, Diane and Dingsøyr, Torgeir and Lindsjorn, Yngve},
    month = mar,
    year = {2022},
    pages = {56},
}

@inproceedings{hukkelbergExploringChallengesIntegrating2019,
    address = {Cham},
    title = {Exploring the {Challenges} of {Integrating} {Data} {Science} {Roles} in {Agile} {Autonomous} {Teams}},
    isbn = {978-3-030-30126-2},
    language = {en},
    booktitle = {Agile {Processes} in {Software} {Engineering} and {Extreme} {Programming} – {Workshops}},
    publisher = {Springer},
    author = {Hukkelberg, Ivar and Berntzen, Marthe},
    editor = {Hoda, Rashina},
    year = {2019},
    pages = {37--45},
}

@book{ulfsnesExploringOrganizationalModels2024,
    title = {Exploring the {Organizational} {Models} for {Data} {Science} in {Agile} {Software} {Development}: {Challenges} and {Strategies} from a {Multi}-{Case} {Study}},
    isbn = {978-0-9981331-7-1},
    shorttitle = {Exploring the {Organizational} {Models} for {Data} {Science} in {Agile} {Software} {Development}},
    language = {eng},
    urldate = {2026-02-26},
    author = {Ulfsnes, Rasmus and Berntzen, Marthe and Moe, Nils Brede and Barbala, Astri and Sporsem, Tor and Stray, Viktoria},
    month = jan,
    year = {2024},
}

@article{giray_software_2021,
	title = {A software engineering perspective on engineering machine learning systems: {State} of the art and challenges},
	volume = {180},
	issn = {01641212},
	shorttitle = {A software engineering perspective on engineering machine learning systems},
	language = {en},
	urldate = {2026-05-06},
	journal = {Journal of Systems and Software},
	author = {Giray, Görkem},
	month = oct,
	year = {2021},
	pages = {111031},
}

@INPROCEEDINGS{amershi_se_ai,
  author={Amershi, Saleema and Begel, Andrew and Bird, Christian and DeLine, Robert and Gall, Harald and Kamar, Ece and Nagappan, Nachiappan and Nushi, Besmira and Zimmermann, Thomas},
  booktitle={2019 IEEE/ACM 41st International Conference on Software Engineering: Software Engineering in Practice (ICSE-SEIP)}, 
  title={Software Engineering for Machine Learning: A Case Study}, 
  year={2019},
  volume={},
  number={},
  pages={291-300}
}

@article{braun_using_2006,
  author  = {Braun, Virginia and Clarke, Victoria},
  title   = {Using thematic analysis in psychology},
  journal = {Qualitative Research in Psychology},
  volume  = {3},
  number  = {2},
  pages   = {77--101},
  year    = {2006}
}

@inproceedings{stray2025human,
author = {Stray, Viktoria and Barbala, Astri and Wivestad, Viggo Tellefsen},
title = {Human{-AI} Collaboration in Software Development: A Mixed-Methods Study of Developers’ Use of {GitHub Copilot and ChatGPT}},
year = {2025},
isbn = {9798400712760},
publisher = {Association for Computing Machinery},
address = {New York, NY, USA},
booktitle = {Proceedings of the 33rd ACM International Conference on the Foundations of Software Engineering},
pages = {1325--1332},
numpages = {8},
series = {FSE Companion '25}
}

@inproceedings{stray2025generative,
  title={Generative {AI and Developer Workflows: How GitHub Copilot and ChatGPT Influence Solo and Pair programming}},
  author={Stray, Viktoria and Moe, Nils Brede and Ganeshan, Nivethika and Kobbenes, Simon},
  booktitle={Proceedings of the 58th Hawaii International Conference on System Sciences},
    doi = {10.24251/HICSS.2025.883},
  year={2025}
}

@article{ravn_team_autonomy,
author = {Ravn, Johan E. and Moe, Nils Brede and Stray, Viktoria and Seim, Eva Amdahl},
title = {Team autonomy and digital transformation: Disruptions and adjustments in a well-established organizational principle},
year = {2022},
issue_date = {Jun 2022},
publisher = {Springer-Verlag},
address = {Berlin, Heidelberg},
volume = {37},
number = {2},
issn = {0951-5666},
url = {https://doi.org/10.1007/s00146-022-01406-1},
doi = {10.1007/s00146-022-01406-1},
journal = {AI Soc.},
month = jun,
pages = {701–710},
numpages = {10}
}

\end{document}